\documentclass[cits]{PoS}
\def\Msol{\mbox{ }M_{\odot}}
\def\Lsol{\mbox{ }L_{\odot}}
\def\Rsol{\mbox{ }R_{\odot}}

\def\Rstar{\mbox{ }R_{\star}}
\def\ergs{\mbox{ erg\,s}^{-1}}

\def\amin{^\prime}
\def\adeg{^{\circ}}
\def\nh{N_{\rm H}}
\def\cmmoinsdeux{\mbox{ cm}^{-2}}
\def\mags{\mbox{ magnitudes}}

\def\igrjstu{\mbox{IGR~J16318$-$4848}}

\title{An INTEGRAL view of High Mass X-ray Binaries : their nature, formation and evolution}

\ShortTitle{An INTEGRAL view of High Mass X-ray Binaries : their nature, formation and evolution}

\author{\speaker{Sylvain Chaty}\thanks{
I would like to thank the organisers for a successfully organized and interesting workshop in the Bibliothèque Nationale de France.
I am grateful to my close collaborators on the study of {\it INTEGRAL} sources: A. Coleiro, P.A. Curran, Q.Z. Liu, I. Negueruela, A. Paizis, L. Pellizza, F. Rahoui, J. Rodriguez, J.A. Tomsick, J.Z. Yan and J.A. Zurita Heras.
This work was supported by the Centre National d'Etudes Spatiales (CNES), based on observations obtained with MINE -- the Multi-wavelength {\it INTEGRAL} NEtwork --.}\\
        Laboratoire AIM (UMR-E 9005 CEA/DSM-CNRS-Université Paris Diderot)
Irfu/Service d'Astrophysique, Centre de Saclay,  Bât. 709, FR-91191 Gif-sur-Yvette Cedex, France\\
        Institut Universitaire de France, 103, bd Saint-Michel 75005 Paris, France \\
        E-mail: \email{chaty@cea.fr}}

\abstract{We describe here the nature, formation and evolution of the supergiant high mass X-ray binary (HMXB) population, i.e. systems accreting the stellar wind of supergiant stars.
There are now many new observations, from the high-energy side (mainly from the {\it INTEGRAL} satellite), complemented by multi-wavelength observations (mainly in the optical, near and mid-infrared from ESO facilities), showing that a new population of supergiant HMXBs has been recently revealed. We report here on the observational facts about the different categories of HMXBs, allowing to build a consistent scenario explaining the various characteristics of these sources, based on models of accretion in these sources (e.g. transitory accretion disc versus clumpy winds).
We also mention new observations suggesting the existence of evolutionary links between Be and stellar wind accreting supergiant X-ray binaries.
}

\FullConference{"An INTEGRAL view of the high-energy sky (the first 10 years)"
9th INTEGRAL Workshop and celebration of the 10th anniversary of the launch,\\
		October 15-19, 2012\\
		Bibliotheque Nationale de France, Paris, France}

\begin{document}

\section{Introduction: different types of X-ray binaries}

Nearly 50 years after the discovery of the first extra-solar X-ray source, Sco\,X-1 \cite{giacconi:1962}, X-ray astronomy has now reached its age of reason, with a plethora of telescopes and satellites covering the whole electromagnetic spectrum, obtaining precious observations of these powerful celestial objects.
High energy binary systems are composed of a compact object -- neutron star (NS) or black hole (BH) -- orbiting and accreting matter from a companion star (see \cite{chaty:2011c} for a review). The companion star is either a low mass star (typically $\sim 1 \Msol$ or less, with a spectral type later than B, called LMXB for ``Low-Mass X-ray Binary''), or a luminous early spectral type OB high mass companion star (typically $> 10 \Msol$, called HMXB for ``High-Mass X-ray Binary''). 300 high energy binary systems are known in our Galaxy: 187 LMXBs and 114 HMXBs (respectively 62\% and 38\% of the total; \cite{liu:2006} \cite{liu:2007}).

Accretion of matter is different for both types of sources. In the case of LMXBs, the small and low mass companion star fills and overflows its Roche lobe, therefore accretion of matter always occurs through the formation of an accretion disc. The compact object can be either a NS or a BH, Sco X-1 falling in the former category.
In the case of HMXBs, accretion can also occur through an accretion disc, for systems in which the companion star overflows its Roche lobe; however this is generally not the case, and there are two alternatives. The first one concerns stars with a circumstellar disc, and it is when the compact object -- on a wide and eccentric orbit -- crosses this disc, that accretion periodically occurs (case of HMXBs containing a main sequence early spectral type Be III/IV/V star, rapidly rotating, called in the following BeHMXBs). The second case is when the massive star ejects a slow and dense wind radially outflowing from the equator, and the compact object directly accretes the stellar wind through e.g. Bondy-Hoyle-Littleton processes (case of HMXBs containing a supergiant I/II star, called sgHMXBs).

%

\section{The $\gamma$-ray sky seen by the {\it INTEGRAL} observatory}
The {\it INTEGRAL} observatory is an ESA satellite launched on 17 October
2002 by a PROTON rocket on an eccentric orbit. It hosts 4
instruments: 2 $\gamma$-ray coded-mask telescopes -- the imager IBIS
and the spectro-imager SPI, observing in the range 10 keV-10 MeV, with
a resolution of $12\amin$ and a field-of-view of $19\adeg$ --, a
coded-mask telescope JEM-X (3-100 keV), and an optical telescope
(OMC).
%
%
The $\gamma$-ray sky seen by {\it INTEGRAL} is very rich, since 723 sources
have been detected by {\it INTEGRAL}, reported in the $4^{th}$ IBIS/ISGRI soft
$\gamma$-ray catalogue, spanning nearly 7 years of observations in the 17-100
keV domain \cite{bird:2010}\footnote{See an up-to-date list at {\em http://irfu.cea.fr/Sap/IGR-Sources/}, maintained by J. Rodriguez and A. Bodaghee}.  
%
Among these sources, there are 185 X-ray binaries (representing 26\% of the whole sample of sources detected by {\it INTEGRAL}, called ``IGRs'' in the following), 255 Active Galactic Nuclei (35\%), 35 Cataclysmic Variables (5\%), and $\sim 30$ sources of other type (4\%): 15 SNRs, 4 Globular Clusters, 3 Soft $\gamma$-ray Repeaters, 2 $\gamma$-ray bursts, etc. 
215 objects still remain unidentified (30\%).
X-ray binaries are separated into 95 LMXBs and 90 HMXBs, each category representing $\sim 13$\% of IGRs. Among identified HMXBs, there are 24 BeHMXBs and 19 sgHMXBs, representing respectively 31\% and 24\% of HMXBs.

It is interesting to follow the evolution of the ratio between BeHMXBs and sgHMXBs. During the pre-{\it INTEGRAL} era, HMXBs were mostly BeHMXBs. For instance, in the catalogue of 130 HMXBs by \cite{liu:2000}, there were 54 BeHMXBs and 5 sgHMXBs (respectively 42\% and 4\% of the total number of HMXBs). Then, the situation changed drastically with the first HMXBs identified by {\it INTEGRAL}: in the catalogue of 114 HMXBs (+128 in the Magellanic Clouds) of \cite{liu:2006}, 60\% of the total number of HMXBs were firmly identified as BeHMXBs and 32\% as sgHMXBs. Therefore, while the ratio of BeHMXBs/HMXBs increased by a factor of 1.5 only, the sgHMXBs/HMXBs ratio increased by a larger factor of 8.
The ISGRI energy range ($> 20$\,keV), immune to the absorption that prevented the discovery of intrinsically absorbed sources by earlier soft X-ray telescopes, allowed us to go from a study of individual sgHMXBs (such as GX\,301-2, 4U\,1700-377, Vela\,X-1, etc.) to a comprehensive study of the characteristics of a whole population of HMXBs\footnote{See for instance \cite{coleiro:2013} and also Coleiro and Chaty, this volume, on the Galactic distribution of HMXBs.}...



\begin{figure}
\centerline{\includegraphics[width=7.cm]{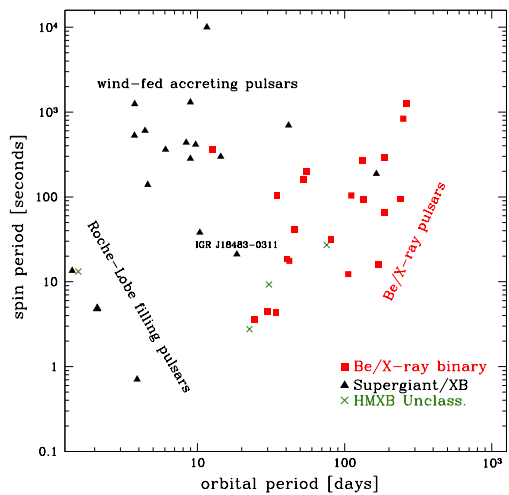}}
\caption{\label{corbet} Corbet diagram (NS $P_\mathrm{spin}$ versus system $P_\mathrm{orb}$), showing the three HMXB populations: {\it i.} systems hosting Be stars (BeHMXBs), {\it ii.} supergiant stars overflowing their Roche lobe, and {\it iii.} systems accreting the stellar wind of supergiant stars (sgHMXBs). The distinct location of these three types of HMXBs results from the interaction and equilibrium between accretion of matter and NS spin. 
The ``misplaced'' SFXT IGR\,J18483-0311 is indicated (credit J.A. Zurita Heras).}
\end{figure}

The most important result of {\it INTEGRAL} to date is the discovery of many new high energy sources -- concentrated in the Galactic plane, mainly towards tangential directions of Galactic arms, rich in star forming regions --, exhibiting common characteristics which previously had rarely been seen (see e.g. \cite{chaty:2011c}). Many of them are HMXBs hosting a NS orbiting an OB companion, in most cases a supergiant star. Nearly all the {\it INTEGRAL} HMXBs for which both $P_\mathrm{spin}$ and $P_\mathrm{orb}$ have been measured are located in the upper part of the Corbet diagram (see Figure \ref{corbet}). They are X-ray wind-accreting pulsars typical of sgHMXBs, with longer pulsation periods and higher absorption compared to previously known sgHMXBs. 
They divide into two classes: some are very obscured, exhibiting a huge intrinsic and local extinction, -- the most extreme example being the highly absorbed source IGR~J16318-4848 \cite{filliatre:2004} --, and the others are sgHMXBs exhibiting fast and transient outbursts -- an unusual characteristic among HMXBs --.  These are therefore called Supergiant Fast X-ray Transients (SFXTs, \cite{negueruela:2006a}),
with IGR~J17544-2619 being their archetype \cite{pellizza:2006}.

\subsection{Obscured HMXBs: IGR~J16318-4848, an extreme case}

  IGR~J16318-4848 was the first source discovered by IBIS/ISGRI on
  {\it INTEGRAL} on 29 January 2003 \cite{courvoisier:2003}. {\it XMM-Newton} observations revealed a
  comptonised spectrum exhibiting an unusually high level of
  absorption: $\nh \sim 1.84 \times 10^{24} \cmmoinsdeux$
  \cite{matt:2003}.  The accurate localisation by {\it XMM-Newton}
  allowed \cite{filliatre:2004} to rapidly trigger ToO photometric and
  spectroscopic observations in optical/infrared, leading to the
  confirmation of the optical counterpart \cite{walter:2003} and to
  the discovery of an extremely bright infrared source (B\,$>25.4\pm1$; I\,$=16.05\pm0.54$; J\,$= 10.33\pm 0.14$; H\,$=8.33\pm 0.10$ and K$_{\mathrm s} =7.20 \pm 0.05 \mags$) \cite{filliatre:2004}. This source exhibits an unusually strong intrinsic absorption in the optical ($A_{\mathrm V} = 17.4\mags$), 100 times stronger than the interstellar absorption along the line of sight ($A_{\mathrm V} = 11.4 \mags$), but still 100 times lower than the absorption in X-rays.

All strong emission lines originate from a highly complex, stratified
circumstellar environment of various densities and temperatures,
suggesting the presence of an envelope and strong stellar outflow/wind
responsible for the absorption. Only luminous early spectral type stars show such extreme environments, and \cite{filliatre:2004} concluded that IGR~J16318-4848 was an unusual HMXB hosting a sgB[e] with characteristic luminosity $10^6 \Lsol$, mass $30 \Msol$, radius $20 \Rsol$ and temperature $T=20250$\,K, located at a distance between 1 and 6 kpc.

\begin{figure}
  \centerline{\includegraphics[width=10.6cm]{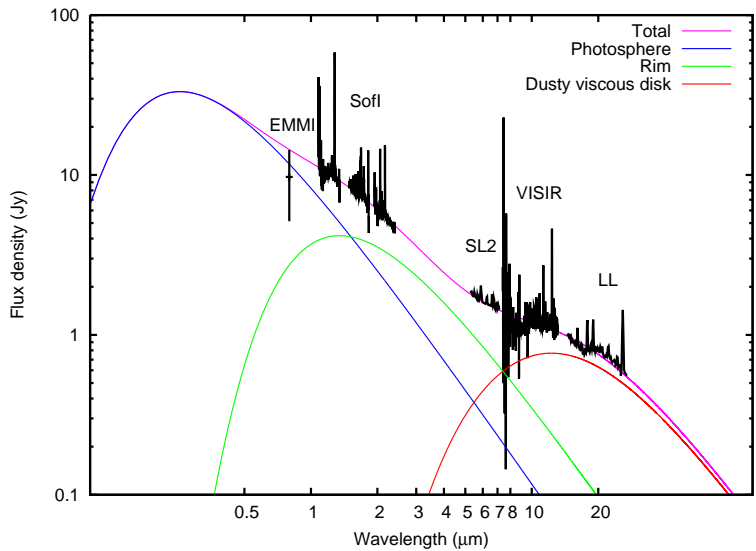}}
\caption{\label{figure:16318}
Broadband near- to mid-infrared ESO/NTT+VISIR and {\it Spitzer} reddened spectrum of
$\igrjstu$, from 0.9 to 35 $\mu$m \cite{chaty:2012}.
}
\end{figure}

Recent MIR spectroscopic observations with VISIR at the VLT and {\it Spitzer} showed that the source was exhibiting strong emission lines of H, He, Ne, PAH, Si, proving that the extra absorbing component was made of dust and cold gas \cite{chaty:2012}. By fitting the optical to MIR spectra with a more sophisticated aspheric disc model developped for HAeBe objets, and adapted to sgB[e] stars, they showed that the supergiant star was surrounded by a hot rim of dust at $5500$\,K, with a warm dust component at $900$\,K around it (see Figure \ref{figure:16318}, \cite{chaty:2012}). This result suggests that this dense and absorbing circumstellar material envelope enshrouds the whole binary system, like a cocoon of dust (see Figure \ref{figure:obscured-sfxt}, left panel).

\subsection{Supergiant Fast X-ray Transients}


SFXTs constitute a new class of $\sim 12$ sources identified among the recently discovered IGRs. They are HMXBs hosting NS orbiting sgOB companion stars, exhibiting peculiar characteristics compared to ``classical'' HMXBs: rapid outbursts lasting only for hours, faint quiescent emission, and high energy spectra requiring a BH or NS accretor. The flares rise in tens of minutes, last for $\sim$ 1 hour, their frequency is $\sim7$\,days, and their luminosity $L_X$ reaches $\sim 10^{36} \ergs$ at the outburst peak.
SFXTs can be divided in two groups, according to the duration and frequency of their outbursts, and their $\frac{L_\mathrm{max}}{L_\mathrm{min}}$ ratio. ``Classical'' SFXTs exhibit a very low quiescence $L_X$ and a high variability, while ``intermediate'' SFXTs exhibit a higher average $L_X$, a lower $\frac{L_\mathrm{max}}{L_\mathrm{min}}$ and a smaller variability factor, with longer flares.  SFXTs might appear like persistent sgHMXBs with $<L_X>$ below the canonical value of $\sim 10^{36} \ergs$, and superimposed flares \cite{walter:2007}, but there might be some observational bias in these general characteristics, therefore the distinction between SFXTs and sgHMXBs is not yet well defined.
While the typical hard X-ray variability factor (the ratio between outburst flux and deep quiescence) is less than 20 in classical/absorbed systems, it is higher than 100, reaching $10^4$ for some SFXTs (some sources can exhibit flares in a few minutes, like for instance XTE\,J1739-302 \& IGR\,J17544-2619, see e.g. \cite{pellizza:2006}).

To explain the emission of SFXTs in the context of sgHMXBs, \cite{negueruela:2008} invoked the existence of two zones in the stellar wind created by the supergiant star, of high and low clump density\footnote{Other models describe these accretion processes, like the formation of transient accretion discs \cite{ruffert:1996, ducci:2010} and the accretion with centrifugal/magnetic barriers \cite{bozzo:2008}.}. This would naturally explain both the X-ray lightcurves, each outburst being due to the accretion of single clumps, and the smooth transition between sgHMXBs and SFXTs, and the existence of intermediate systems; the main difference between classical sgHMXBs and SFXTs being in this scenario the NS orbital radius.
Indeed, a basic model of porous wind with macro-clumping \cite{negueruela:2008} predicts a substantial change in the properties of the wind ``seen by the NS'' at a distance $r \sim 2 \Rstar$, where we stop seeing persistent X-ray sources. There are 2-regimes:
at $r < 2 \Rstar$ the NS sees a large number of clumps, embedded in a quasi-continuous wind; and at $r > 2 \Rstar$ the density of clumps is so low that the NS is effectively orbiting in an empty space.
NS in classical sgHMXBs can only lie within $2 \Rstar$ of the companion star.

\subsubsection{What is the origin of ``misplaced'' sgHMXBs?}

\cite{liu:2011} noted that there are two ``misplaced'' SFXTs in the Corbet diagram: IGR\,J11215-5952 (a NS orbiting a B1 Ia star, \cite{negueruela:2005b}) and IGR\,J18483-0311 (\cite{rahoui:2008a}; see Figure \ref{corbet}). According to \cite{liu:2011}, these 2 SFXTs can not have evolved from normal main sequence OB-type stars, since {\it i.} their NS are not spinning at the equilibrium spin period $P_\mathrm{eq}$ of O\,V stars, {\it ii.} they have not been able to spin up after reaching $P_\mathrm{eq}$, and {\it iii.} their NS have not yet reached $P_\mathrm{eq}$ of sgHMXBs (see e.g. \cite{waters:1989}). They must therefore be the descendants of BeHMXBs (i.e. hosting O-type emission line stars), after the NS has reached $P_\mathrm{eq}$, suggesting that some HMXBs exhibit two periods of accretion, the first one as BeHMXB and the second one as sgHMXB \cite{liu:2011}. This important result suggests that there must be many more such intermediate SFXTs.


\subsection{A scenario towards the Grand Unification of sgHMXBs}

\begin{figure}
\includegraphics[height=.235\textheight,angle=0]{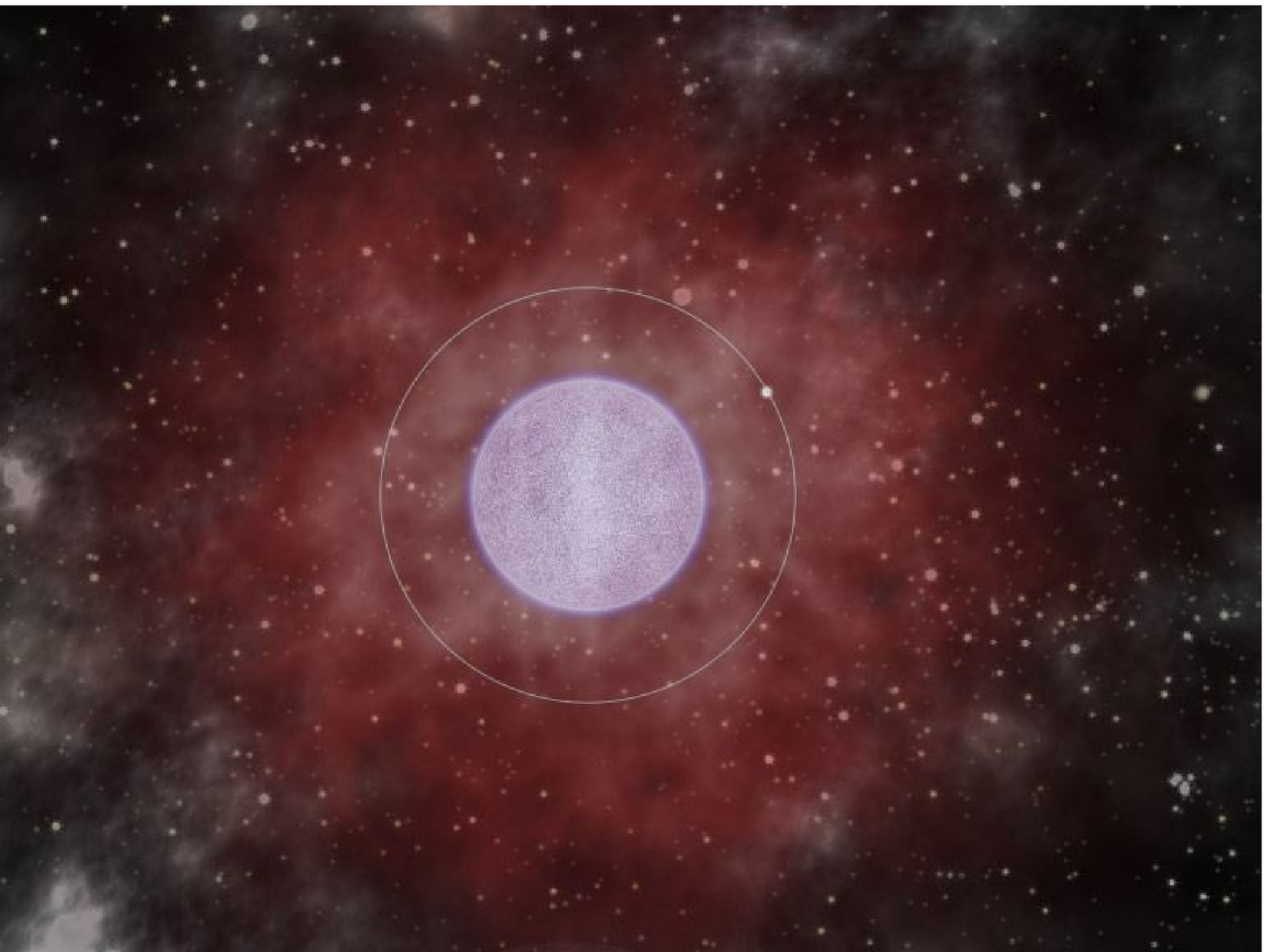}
\includegraphics[height=.235\textheight,angle=0]{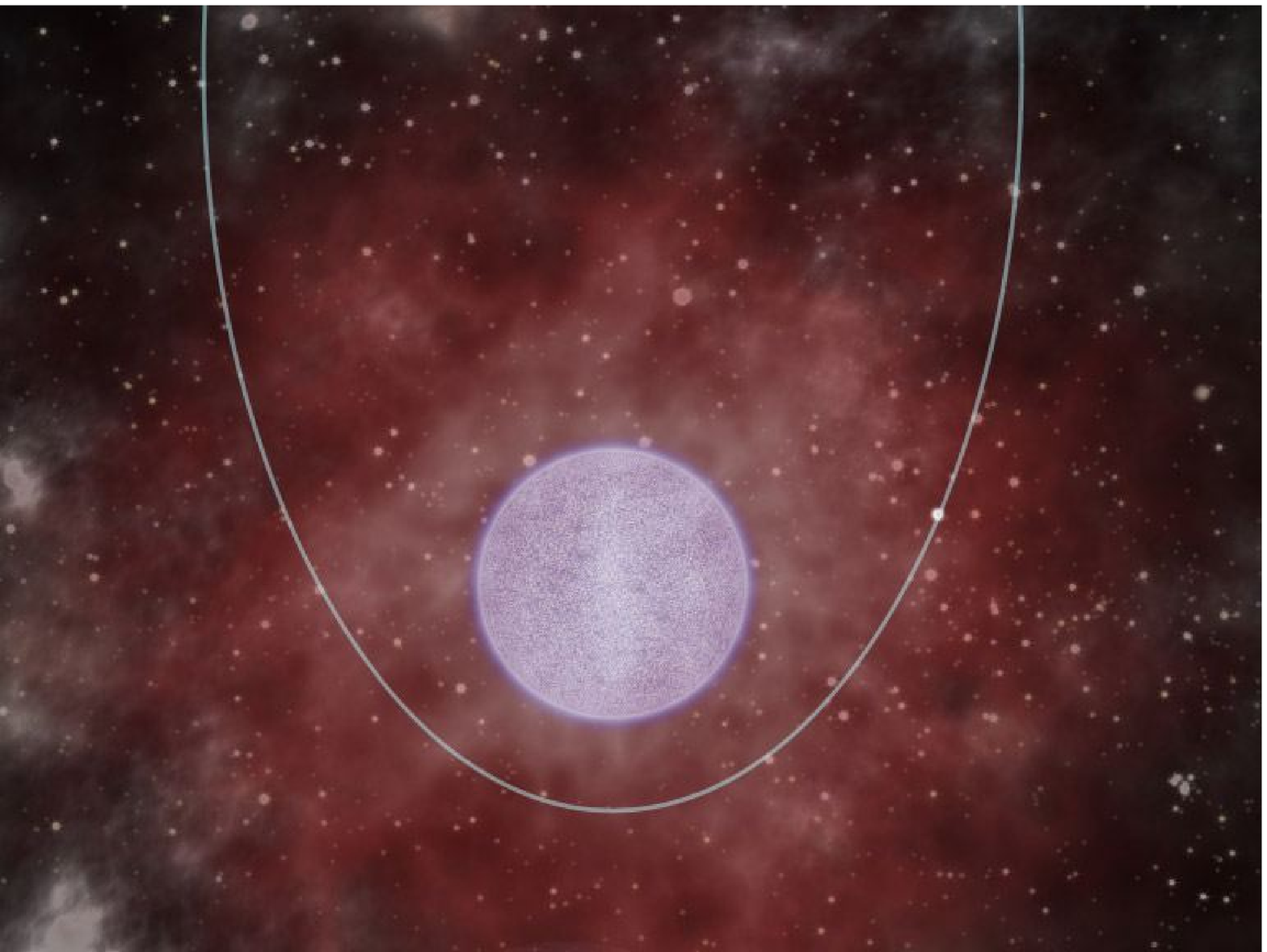}
\caption{\label{figure:obscured-sfxt}
  Scenario illustrating two possible configurations of {\it INTEGRAL} 
  sources: a NS orbiting a supergiant
  star on a circular orbit (left image); and on an eccentric orbit
  (right image), accreting from the clumpy stellar wind of the
  supergiant.  The accretion of matter is persistent in the case of
  the obscured sources, as in the left image, where the compact object
  orbits inside the cocoon of dust enshrouding the whole system. On
  the other hand, the accretion is intermittent in the case of SFXTs,
  which might correspond to a compact object on an eccentric orbit, as
  in the right image.  A 3D animation of these sources is available on the
  website: {\em http://www.aim.univ-paris7.fr/CHATY/Research/hidden.html}
}
\end{figure}

In view of the results described above, there seems to exist a continuous trend, with the observed division between classical sgHMBs and SFXTs being naturally explained by simple geometrical differences in the orbital configurations (see e.g. \cite{chaty:2011c}):

\begin{description}

\item [Classical (or obscured/absorbed) sgHMXBs:]
  These systems (like IGR\,J16318-4848) host a NS on  a short and circular orbit at a few stellar radii only from the star, inside the zone of stellar wind high clump density ($R_\mathrm{orb} \sim 2\Rstar$). It is constantly orbiting inside a cocoon of dust and/or cold gas, probably created by the companion star itself, therefore creating a persistent and luminous X-ray emission. The cocoon, with an extension of $\sim 10 \Rstar = 1$\,a.u., is enshrouding the whole binary system (see Figure \ref{figure:obscured-sfxt}, left panel).

\item [Intermediate SFXT systems:]
In these systems (such as IGR\,J18483-0311, $P_\mathrm{orb} = 18.5$\,days), the NS orbits on a short and circular/eccentric orbit outside the zone of high clump density, and it is only when the NS penetrates inside the narrow transition zone between high and low clump density, close to the supergiant star, that accretion takes place, and that X-ray emission arises, with possible periodic outbursts.

\item [Classical SFXTs:]
In these systems (such as XTE\,J1739-302, $P_\mathrm{orb} = 50$\,days), the NS orbits outside the high density zone, on a large and eccentric orbit around the supergiant star, and exhibits some recurrent and short transient X-ray flares, when it comes close to the star, and accretes from clumps of matter coming from the wind of the supergiant.  Because it is passing through more diluted medium, the $\frac{Lmax}{Lmin}$ ratio is higher and the quiescence lasts for longer time for classical SFXTs compared to intermediate SFXTs (see Figure \ref{figure:obscured-sfxt}, right panel).

\end{description}

Although this scenario seems to describe quite well the characteristics
currently seen in sgHMXBs, we still need to identify the nature of
many more sgHMXBs to confirm it, and in particular $P_\mathrm{orb}$ and the dependance of the column density with the orbital phase of the binary system.

\section{Conclusions and perspectives}

Let us first recall the {\it INTEGRAL} legacy:

\begin{itemize}

\item The {\it INTEGRAL} satellite has quadrupled the total number of known Galactic sgHMXBs, constituted of a NS orbiting a supergiant star. Most of the new sources are slow and absorbed X-ray pulsars, exhibiting a large $\nh$ and long $P_\mathrm{spin}$ ($\sim1$\,ks).

\item The {\it INTEGRAL} satellite has revealed the existence in our Galaxy of two previously hidden populations of high-energy binary systems.
Firstly, the SFXTs, exhibiting brief and intense X-ray flares -- with a peak flux of 1 Crab during 1--100s every $\sim 100$\,days --, which can be explained by accretion through clumpy winds.
Secondly, a previously hidden population of obscured and persistent sgHMXBs, composed of supergiant companion stars exhibiting a strong intrinsic absorption and long $P_\mathrm{spin}$, with the NS deeply embedded in the dense stellar wind, forming a dust cocoon enshrouding the whole binary system.

\end{itemize}

Apart from these observational facts, has {\it INTEGRAL} allowed us to better understand sgHMXBs and other populations of HMXBs? Do we better apprehend the accretion processes in HMXBs in general, and in sgHMXBs in particular, and what makes the fast transient flares so special, in the context of the clumpy wind model, and of the formation of transient accretion discs? 
In summary, has the increased population of supergiant HMXBs allowed us a better knowledge of these sources, compared to the ones that were already known before the launch of {\it INTEGRAL}? The answer to all these questions is probably {\it ``not yet''}, however we now have in hand more sources, and therefore more constraints to play with.
Studying these populations will provide a better understanding of the formation and evolution of short-living HMXBs, and study accretion processes. Furthermore, stellar population models now have to take these objects into account, to assess a realistic number of
high-energy binary systems in our Galaxy. 



\end{document}